\documentclass[12pt]{article}
\usepackage{putex}
\usepackage{graphicx}
\begin{document}

\preprint{hep-th/0411257 \\ PUPT-2144}

\institution{PU}{Joseph Henry Laboratories, Princeton University, Princeton, NJ 08544}

\title{The Gregory-Laflamme instability for the D2-D0 bound state}

\authors{Steven S. Gubser\footnote{e-mail: {\tt ssgubser@Princeton.EDU}}}

\abstract{The D2-D0 bound state exhibits a Gregory-Laflamme instability when it is sufficiently non-extremal.  If there are no D0-branes, the requisite non-extremality is finite.  When most of the extremal mass comes from D0-branes, the requisite non-extremality is very small.  The location of the threshold for the instability is determined using a local thermodynamic analysis which is then checked against a numerical analysis of the linearized equations of motion.  The thermodynamic analysis reveals an instability of non-commutative field theory at finite temperature, which may occur only at very long wavelengths as the decoupling limit is approached.}

\date{November 2004}

\maketitle

\tableofcontents

\section{Introduction}
\label{INTRODUCTION}

The tension of the D2-D0 bound state is 
 \eqn{BPStension}{
  \tau_{D2-D0} = \sqrt{N_2^2 \tau_{D2}^2 + \rho_0^2 \tau_{D0}^2} 
   \,.
 }
Here $\tau_{D2}$ and $\tau_{D0}$ are the D2-brane tension and D0-brane mass, respectively.  $N_2$ is the number of D2-branes, and $\rho_0 = N_0/V_2$ is the number of D0-branes per unit world-volume area.  We are considering the bound state at extremality in asymptotically flat ten-dimensional spacetime.

The D2-D0 bound state is supersymmetric and therefore stable.  Without referring to supersymmetry, we can see from \BPStension\ that the bound state is stable against infrared perturbations that make the number density of D0-branes non-uniform: the key fact is that $\tau_{D2-D0}$ is a convex function of $\rho_0$.  More precisely, suppose we make a long-wavelength perturbation, $\rho_0 \to \rho_0 + \delta\rho_0$.  Then the change in the total mass of the bound state is
 \eqn{deltaM}{
  \delta M \approx \int d^2 \xi \, \sqrt{g} \left(
   \sqrt{N_2^2 \tau_{D2}^2 + (\rho_0 + \delta\rho_0)^2 \tau_{D0}^2} - 
   \sqrt{N_2^2 \tau_{D2}^2 + \rho_0^2 \tau_{D0}^2}
   \right) \,.
 }
The right hand side is always positive, provided $\int d^2 \sigma \sqrt{g} \, \delta\rho_0 = 0$, precisely because the tension formula \BPStension\ is convex in $\rho_2$.  The equality in \deltaM\ is approximate because there should also be contributions expressed in terms of spatial derivatives of $\delta\rho_0$.  The point of considering only infrared fluctuations is to suppress these contributions.  Within this framework, we have reproduced the expected conclusion that the D2-D0 bound state is stable against local rearrangements of D0-brane charge.

The purpose of this paper is essentially to extend the above analysis to a non-extremal D2-D0 bound state, where it is not so obvious whether there should be an instability.  All calculations will be done using the supergravity description of the bound state, so the quantitative results are limited to supergravity's regime of validity.  However, I expect that the following qualitative picture holds more generally for D2-D0 bound states without angular momentum:
 \begin{itemize}
  \item Far enough from extremality, there is an instability.  Except in special limits, the instability disappears when the mass above extremality is some $O(1)$ multiple of the extremal mass.
  \item When D0-branes make up most of the mass of the bound state, the instability persists close to extremality: the aforementioned $O(1)$ factor becomes small.
 \end{itemize}
One of our main tools is a notion of local thermodynamic stability.  Briefly, at finite temperature, we may independently rearrange D0-brane charge and entropy, so to redo the above calculation properly, we should demand that the energy is a convex function of them both in order to avoid an instability.  More precisely, the Hessian matrix of second derivatives of the energy with respect to the D0-brane charge and the entropy should be positive definite to avoid an instability.

In \cite{gmOne,gmTwo} it was conjectured that for systems with translational symmetry and infinite extent, a Gregory-Laflamme (GL) instability \cite{glOne,glTwo} arises precisely when the system has a thermodynamic instability---meaning that the Hessian matrix of second derivatives of the mass with respect to entropy and any charges which are capable of being redistributed over the directions in which the GL instability is supposed to occur.  I will refer to this as the correlated stability conjecture (CSC).  A heuristic justification of the CSC is that if one writes entropy in terms of the mass and charges, then a positive eigenvalue of the Hessian matrix identifies a direction of evolution in which entropy can increase while total mass and charge remain constant.  The idea that the GL instability is associated with entropy increase was already present in the original papers \cite{glOne,glTwo}.  The CSC makes this idea more systematic by equating a local thermodynamic instability to a perturbative dynamical instability.

For classical systems with horizons of finite extent (e.g.~a Schwarzschild black hole, or a black string wrapped on a circle), there can be no conclusive argument for the existence of a horizon instability based solely on thermodynamics.  The reason is {\it not} that infinite volume is necessary for a thermodynamic limit to be taken: with the freedom to rescale the Planck scale, any classical horizon can be regarded as large.  Instead, the reason is that the wavelength for the instability may be larger than the system size.  The generic expectation (with interesting exceptions near the boundary of stability) is that the wavelength of the GL instability is of the same order as the horizon radius.  The claim, then, for finite size systems, is that the existence or non-existence of a GL instability is driven entirely by the competition between thermodynamic and finite-size effects.  From this point of view, the Schwarzschild black hole in four-dimensions is stable only because of $O(1)$ effects: it's just a bit too small for its natural GL instability to fit on its horizon. 

A reason to be particularly interested in the GL instability for the {\it near}-extremal D2-D0 bound state is that it interpolates between {\it near}-extremal D2-branes, for which it is known there is no instability \cite{gaUniv} and {\it near}-extremal smeared D0-branes, for which it has been argued that there is a GL instability \cite{RossBostock,AharonyEtAlOne,Harmark:2004ws}.  The interpolation itself is interesting: the dynamics of the bound state has been argued to correspond to non-commutative field theory (NCFT) \cite{swNonCom}.  More precisely, a particular limit, in which the energy scale of allowed processes is lowered at the same time as the density of D0-branes is increased, results in a decoupling of closed strings and excited open strings from the non-commutative dynamics of massless open strings.

Because the NCFT limit corresponds to an arbitrarily large density of D0-branes, the convexity property that ``holds together'' the bound state is extremely weak: inspecting \BPStension\ for large $\rho_0$, we see that it is barely convex at large $\rho_0$.  So we might think that a very slight non-extremality causes an instability.  The question is whether the non-extremal temperature is bigger or smaller than the energy scale of allowed excitations in the NCFT limit.  I will argue that it is smaller.  Should we then conclude that NCFT is unstable at any non-zero temperature?  The instability corresponds to the non-commutativity parameter $\vartheta_{\mu\nu}$ becoming inhomogeneous.  It's possible that the wavelength of this instability diverges in the NCFT limit.  In other words, NCFT might avoid the instability by pushing it to infinite wavelength.  This scenario seems likely based on decoupling arguments, but I do not have an explicit computation of the wavelength.

In section~\ref{SUPERGRAVITY} I briefly present the supergravity background describing the D2-D0 bound state.  In section~\ref{CSC} I use the CSC to predict for which range of parameters a GL instability should occur.  In section~\ref{NONCOMMUTATIVE} I argue that the transition temperature to a GL instability vanishes as one approaches the NCFT limit.  In section~\ref{INSTABILITY} I perform a numerical analysis on the equations of motion linearized around the D2-D0 background which confirms the predictions of the CSC.  Section~\ref{CONCLUDE} comprises some concluding remarks.

\section{Classical action, solutions, and thermodynamics}
\label{SUPERGRAVITY}

The part of the classical type~IIA supergravity action relevant for the considerations of this paper is
 \eqn{IIAaction}{
   S &= {1 \over 2\kappa^2} \int d^{10} x \, \sqrt{G}
   \left[ e^{-2\phi} 
    \left( R + 4(\partial\phi)^2 - {1 \over 2} H_3^2 \right) - 
    {1 \over 2} F_2^2 - {1 \over 2} \tilde{F}_4^2 \right]  \cr
   &\qquad\qquad\hbox{with $\tilde{F}_4 = F_4 + A_1 \wedge H_3$} \,.
 }
The non-extremal D2-D0 bound state corresponds to the following supergravity background:
 \eqn{DzeroDtwo}{
  ds_{str}^2 &= H^{-1/2} (-h dt^2 + D (dx_1^2 + dx_2^2)) + 
   H^{1/2} \left( {1 \over h} dr^2 + r^2 d\Omega_6^2 \right)  \cr
    A_1 &= \coth\alpha \sin\theta
      \left( 1 - {1 \over H} \right) dt  \cr
    A_3 &= \coth\alpha\sec\theta \left( 1 - {D \over H} \right)
      dt \wedge dx^1 \wedge dx^2 \cr
    e^{2\phi} &= H^{1/2} D \qquad 
    B_2 = \tan\theta \left( 1 - {D \over H} \right) 
     dx^1 \wedge dx^2 \cr
    H &= 1 + {r_0^5 \sinh^2 \alpha \over r^5}\qquad
    D = {1 \over H^{-1} \sin^2 \theta + \cos^2 \theta}\qquad
    h = 1 - {r_0^5 \over r^5} \,.
 }
The thermodynamic quantities of interest are
 \eqn{thermo}{
  M &= {V_2 \Omega_6 \over 2\kappa^2}
    r_0^5 (6 + 5 \sinh^2\alpha)  \cr
  T &= {5 \over 4\pi r_0 \cosh\alpha} \qquad
    S = {2\pi V_2 \Omega_6 \over \kappa^2} r_0^6 \cosh\alpha  \cr
  \mu_2 &= \mu \cos\theta \qquad Q_2 = Q \cos\theta \qquad 
    \mu_0 = \mu \sin\theta \qquad Q_0 = Q \sin\theta  \cr
  \mu &= \tanh\alpha \qquad 
    Q = {5 V_2 \Omega_6 \over 2\kappa^2} r_0^5 
     \sinh\alpha\cosh\alpha \,,
 }
where $\Omega_6 = {16 \over 15} \pi^3$ is the volume of a unit $S^6$.

Solutions of the type described in \DzeroDtwo\ are well known: see for example \cite{HarmarkObers,Harmark} for substantial generalizations of it, for references to earlier literature, and for some description of how such solutions can be obtained from simpler ones using rotations and T-duality.

\section{Consequences of the correlated stability conjecture}
\label{CSC}

To apply the correlated stability conjecture (CSC), we need to know what extensive thermodynamic quantities are capable of being locally redistributed.  For the non-extremal D2-D0 bound state, the D2-brane charge cannot be redistributed because D2-branes can't break.  But the D0-brane charge can be redistributed.  Also, entropy can be redistributed by making the horizon wavy.  So to determine local thermodynamic stability, one should consider the Hessian of $M$ with respect to $S$ and $Q_0$, holding $Q_2$ fixed.

\begin{figure}
  \centerline{\includegraphics[width=3in]{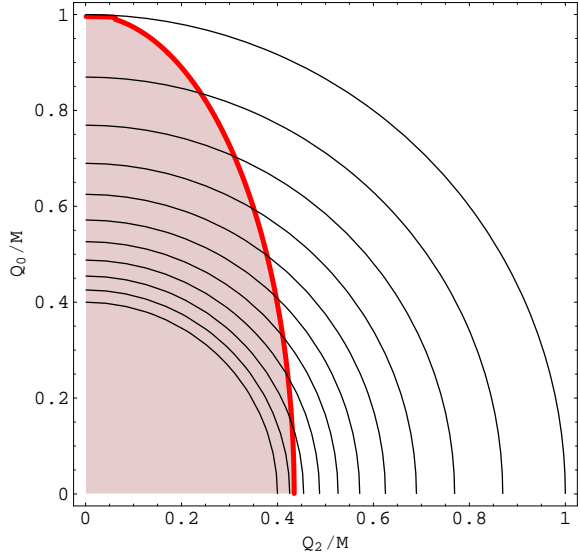}}
  \caption{The thick curve indicates the boundary of stability.  The shaded region is the unstable region.  The thin curves show constant $M$ for fixed $Q = \sqrt{Q_2^2+Q_0^2}$: the outermost of these curves is at $M=Q$, the next is at $M = 1.15 Q$, and so on with even spacing continuing to $M = 2.5 Q$.  The vertical axis and horizontal axis show, respectively, the fraction of the mass that comes from D0-branes and D2-branes; the rest of the mass is non-extremality.}\label{figA}
 \end{figure}

In \DzeroDtwo, $M$ is not expressed as a function of $S$, $Q_2$, and $Q_0$; instead, these four quantities are expressed in terms of $r_0$, $\alpha$, and $\theta$.  So it's useful to recall a fact from multi-variable calculus: if there is a smooth, invertible relationship between $n$ variables $Q_i$ and $n$ other variables $q_i$, and $M$ is known as a smooth function of the $q_i$, then
 \eqn{NiceFact}{
  \left( {\partial M \over \partial Q_i} \right)_{Q_j} = 
   {\partial(Q_1,\ldots,\hat{Q}_i,M,\ldots, Q_n)/
     \partial(q_1,\ldots,q_n) \over
    \partial(Q_1,\ldots,Q_n)/\partial(q_1,\ldots,q_n)} \,,
 }
where the denominator is the Jacobian, $\det(\partial Q_i/\partial q_j)$, and the hat notation in the numerator is meant to indicate replacing $Q_i$ with $M$.  Second derivatives of $M$ can be taken by plugging a first derivative back into \NiceFact\ in place of $M$.  Actually, the thermodynamic dual quantities $T$, $\mu_2$, and $\mu_0$ should be precisely the first derivatives of $M$ with respect to $S$, $Q_2$, and $Q_0$ because of the first law of thermodynamics:
 \eqn{FirstLaw}{
  dE = TdS + \mu_2 dQ_2 + \mu_0 dQ_0 \,.
 }
This consistency check was already made in \cite{HarmarkObers}.  The Hessian may be expressed as $H = \partial(T,\mu_0)/\partial(S,Q_0)$, and the right hand side may be evaluated by repeated use of \NiceFact. One finds
 \eqn{DetH}{
  \det H = {\kappa^4 \sech^4\alpha \over 8 \pi^2 \Omega_6^2
   r_0^{12} V_2^2 (5+7\cosh 2\alpha)} (-7-3\cos 2\theta + 
    6 \cos^2\theta \cosh 2\alpha) \,.
 }
The boundary of local thermodynamic stability is where $\det H = 0$.  (More properly, it is the boundary of the region where all eigenvalues of $H$ are positive, but it turns out in this case that $H$ has at most one negative eigenvalue).  Starting from \DetH\ one can easily demonstrate that $\det H = 0$ is equivalent to
 \eqn{FSC}{
  \csch\alpha = \sqrt{3} \cos\theta \,.
 }
This marginal stability condition can be recast in terms of the potentials $\mu_0$ and $\mu_2$ appearing in \thermo\ as
 \eqn{FSCagain}{
  \mu_0^2 + 4 \mu_2^2 = 1 \,.
 }
To understand this result, it is useful to plot $Q_0/M = 5\sin\theta \sinh\alpha \cosh\alpha / (6+5\sinh^2 \alpha)$ against $Q_2/M = 5\cos\theta \sinh\alpha \cosh\alpha / (6+5\sinh^2 \alpha)$, which is done in figure~\ref{figA}.
Given the curve \FSC\ of local thermodynamic stability, the CSC predicts a GL instability precisely when $\alpha < \csch^{-1} (\sqrt{3} \cos\theta)$.  If there are only D0-branes and no D2-branes (that is, $\theta = \pi/2$), then there is supposed to be a GL instability for all finite values of $\alpha$.

\section{The non-commutative field theory limit}
\label{NONCOMMUTATIVE}

Special interest attaches to the low-energy dynamics of massless open strings on the D2-D0 bound state in the limit where most of the mass of the bound state comes from the D0-branes.  This dynamics has been argued to be a non-commutative field theory (NCFT) decoupled from gravity and excited string states\cite{swNonCom}.  Yet we can see from figure~\ref{figA} that the boundary of stability predicted by the CSC goes through precisely the region of interest for this limit: $Q_0/M \approx 1$ while $Q_2/M \ll 1$.  The GL instability predicted by the CSC involves redistributing the D0-branes, which means making the parameter $\vartheta$ that measures non-commutativity a function of space---and of time unless there is an endpoint to the dynamical evolution of the GL instability.  Such a situation would seem to spoil in an important way the decoupling of the NCFT from closed strings: spatially variable $B_2$ is clearly a closed string degree of freedom.

How might the NCFT limit avoid the GL instability?  One possibility hinges on the fact that it is a low-energy limit.  The boundary of stability shown in figure~\ref{figA} defines a critical temperature $T_c$ at any fixed values of $Q_2$ and $Q_0$.  At least near the upper left corner of figure~\ref{figA}, when D0-branes make the dominant contribution to the mass, solutions on the stable side of the curve have temperature $T < T_c$.  If $T_c$ were large compared to the energies permitted in the NCFT limit, then clearly the GL instability would be altogether avoided in the NCFT limit.  Unfortunately, this is not what happens: as we will show in this section, $T_c$ is small compared to the intrinsic energy scale of the NCFT, $1/\sqrt\vartheta$, where $\vartheta$ is the non-commutativity parameter.

There is another possibility: the critical wavelength above which there are unstable modes might diverge in the NCFT limit, for all temperatures which remain finite as compared to the scale $1/\sqrt\vartheta$.  In principle we should be able to check this numerically, but we have not done so.  In section~\ref{RESULTS} we will briefly describe why this is a difficult numerical problem in practice.  Without hard evidence one way or another, the divergent wavelength possibility is the one I regard as the most likely.  If true, it implies that if we compactify the D2-D0 bound state on a torus whose size is held fixed as one takes the NCFT limit, then there is no GL instability in this limit at all.  A third possibility---that NCFT is unstable toward non-uniform non-commutativity parameter at some finite temperature---would simply mean that the NCFT limit is ill-defined.  I would find this hard to understand in light of the very plausible decoupling arguments put forward in \cite{swNonCom}.

Let us now show that the critical temperature $T_c$ is much less than the intrinsic scale $1/\sqrt\vartheta$ in the NCFT limit.  The demonstration mainly consists of unraveling the definitions for $T_c$ and $\vartheta$ and recalling how the NCFT limit is taken.

Following \cite{swNonCom}, consider the D2-D0 bound state in the test brane approximation.  The metric felt by closed strings is the metric $G_{MN}$ appearing in \DzeroDtwo.  In discussions of NCFT this closed string metric is usually denoted $g_{MN}$, while $G_{MN}$ is reserved for the metric felt by open strings.  I will follow this convention in this section.  A useful quantity is
 \eqn{Edef}{
  E_{\alpha\beta} = g_{\alpha\beta} + B_{\alpha\beta} \,,
 }
where indices $\alpha$, $\beta$ denote a pull-back to the $2+1$-dimensional brane world-volume.  The inverse matrix $E^{\alpha\beta}$ may be expanded into a symmetric and anti-symmetric part:
 \eqn{Eexpand}{
  E^{\alpha\beta} = G^{\alpha\beta} + 
   {\vartheta^{\alpha\beta} \over 2\pi\alpha'} \,.
 }
Then $G_{\alpha\beta}$ (defined as the inverse of $G^{\alpha\beta}$) is indeed the metric felt by the open strings: this means, for instance, that massless open strings follow null geodesics of $G_{\alpha\beta}$. It is convenient to change coordinates on the brane to $\tilde{x}^\alpha$ such that $G_{\alpha\beta} = \diag\{-1,1,1\}$.  One finds that $\vartheta^{\alpha\beta}$ measures non-commutativity:  
 \eqn{noncom}{
  [\tilde{x}^\alpha,\tilde{x}^\beta] = i \vartheta^{\alpha\beta} \,.
 }
Operationally, this means that open string scattering amplitudes come with phases like $e^{-{i \over 2} \vartheta^{\alpha\beta} \tilde{p}_\alpha \tilde{q}_\beta}$ where $\tilde{p}$ and $\tilde{q}$ are external momenta---see \cite{swNonCom} for details.  Assuming only space-space non-commutativity, the NCFT limit is $\alpha' \to 0$ with $\vartheta \equiv \vartheta^{12}$ held fixed.  Then $\sqrt\vartheta$ is an intrinsic length scale characterizing the NCFT.

Consider the case where only $B_{12} = -B_{21} = -\tan\theta$ is non-zero.\footnote{The solution as expressed in \DzeroDtwo\ has $B_{12} = 0$ far from the brane and $B_{12} \approx \tan\theta$ at the horizon, where the approximate equality becomes exact in the near-extremal limit.  A global shift, $B_{12} \to B_{12} - B$, is appropriate to compare with the test brane approximation.}  One can easily show that $\tilde{t} = t$, $\tilde{x}^1 = x^1/\cos\theta$, and $\tilde{x}^2 = x^2/\cos\theta$.  Note that this means that the speed of propagation for massless open strings is $\cos\theta$.  When expressed in the $\tilde{x}$ coordinate system, $\vartheta^{12} = -\vartheta^{21} = \vartheta$, where
 \eqn{varthetaDef}{
  \vartheta = 2\pi\alpha' \tan\theta \,.
 }
In the NCFT limit, $\vartheta$ is held fixed while $\alpha' \to 0$, so one needs $\theta \to \pi/2$.

The parameters used to specify the NCFT are $\vartheta$, the string coupling $g_s$, the number $N$ of D2-branes, and the temperature $T$.  Our aim now is to express the condition \FSC\ defining the boundary of stability in terms of these quantities.  First note that $Q/V_2$ is the tension at extremality, so $Q_2/V_2$ is the tension at extremality of the D2-branes only:
 \eqn{QVtension}{
  {Q_2 \over V_2} = {5\Omega_6 \over 2\kappa^2} 
    r_0^5 \sinh\alpha \cosh\alpha \cos\theta = N_2 \tau_{D2}
     = {N_2 \over g_s 4\pi^2 \alpha'^{3/2}} \,.
 }
Using this equation together with the standard relation $2\kappa^2 = (2\pi)^7 g_s^2 \alpha'^4$, one may eliminate $r_0$ in favor of $g_s$, $N_2$, $\alpha$, $\theta$, and $\alpha'$.  Using in addition \varthetaDef, one may express the dimensionless quantity $\sqrt{\vartheta} T$ in terms of the same parameters:
 \eqn{thetaT}{
  \sqrt{\vartheta} T = {5 \over 2^{17/10} 3^{1/5} \pi^{9/10}}
   {1 \over (g_s N_2)^{1/5}} {(\sinh\alpha)^{1/5} \over
    (\cosh\alpha)^{4/5}} (\cos\theta)^{1/5} \sqrt{\tan\theta} \,.
 }
Note that $T$ is an energy scale measured with respect to $t = \tilde{t}$, while $\sqrt\vartheta$ is a length measured with respect to $\tilde{x}^1$ and $\tilde{x}^2$.  The correct dimensionless quantity to compute is $\sqrt\vartheta T/v$, where $v$ is the propagation speed of massless open strings.  But we are using the coordinates $(t,\tilde{x}^1,\tilde{x}^2)$, both to define $\vartheta$ (see \noncom) and to measure $T$.  And in these coordinates, $v=1$.  So $\sqrt\vartheta T$ really is the right dimensionless quantity after all.

Now let us find the value of $\sqrt{\vartheta} T$ corresponding to the boundary of stability: using \FSC\ to eliminate $\alpha$ from \thetaT\ one obtains
 \eqn{thetaTsimp}{
  \sqrt{\vartheta} T_c &= 
   {5 \cdot 3^{1/10} \over 2^{17/10} \pi^{9/10}} 
    {1 \over (g_s N_2)^{1/5}} (\cot\theta)^{3/10} \left[
     1 - {8 \over 5} \cot^2\theta + O(\cot^4\theta) \right]  \cr
   &= {5 \cdot 3^{1/10} \over 2^{17/10} \pi^{9/10}} 
    {1 \over (g_s N_2)^{1/5}} 
     \left( {2\pi\alpha' \over \vartheta} \right)^{3/10} 
     \left[ 1 + O(\alpha'^2/\vartheta^2) \right] \,.
 }
As described near the beginning of this section, $T_c$ is a transition temperature, below which there is a uniform phase, and above which there is a GL instability which makes $\vartheta$ non-uniform and whose endpoint is not known.

From \thetaTsimp\ it is clear that $\sqrt\vartheta T_c \to 0$ in the limit where $\alpha'/\vartheta \to 0$, assuming finite 't Hooft coupling $g_s N_2$.  This seems to invalidate the claim \cite{mRusso,Cai:2000hn} that NCFT has the same large $N$ thermodynamics as the corresponding commutative theory: introducing an instability for all $T > 0$ is a pretty radical change!  But if the divergent wavelengths hypothesis described near the beginning of this section is correct, the apparent discrepancy is merely an order of limits issue: the claims of \cite{mRusso,Cai:2000hn} would follow if we take the NCFT limit on a torus of fixed size and then make the torus large, whereas to keep the instability in the theory one must keep the torus larger than the critical wavelength as one approaches the NCFT limit.

\section{Perturbing the D2-D0 bound state}
\label{INSTABILITY}

Although well-motivated and well-checked for a wide class of examples, the CSC is not proven in full generality.  I therefore aim to demonstrate explicitly that when the CSC says there should be a GL instability for the D2-D0 bound state, there is a stationary perturbation of the supergravity solution \DzeroDtwo\ that is non-uniform in the coordinate $x^1$.  It is generally understood (see for example \cite{Reall}) that if such a perturbation exists at a finite wavelength, then similar perturbations with longer wavelengths will be unstable.

Obtaining the explicit form of the stationary perturbation requires numerics.  The supergravity action and background, \IIAaction\ and \DzeroDtwo, are somewhat complicated, and the perturbation equations are remarkably complicated.  Therefore I will only present a summary discussion in this paper.  Briefly, in section~\ref{REDUCE} the problem is reduced to two dimensions; then in section~\ref{SETUP} it is explained how one extracts from the two-dimensional lagrangian and constraints a well-posed initial value problem.  Finally, in section~\ref{RESULTS} I describe the results of numerics.

\subsection{Reducing to a two-dimensional lagrangian}
\label{REDUCE}

I will assume that this perturbation is an $s$-wave with respect to the $S^6$: it depends only on $x^1$ and $r$.  Therefore it is natural to start by making a Kaluza-Klein reduction on the $S^6$, $t$, and $x^2$ directions.  The result is an effective lagrangian for several scalars, equation \eno{RLagain}, together with constraints from the two-dimensional Einstein equations described in equation \eno{EEqs}.

The ansatz may be described as follows:
 \eqn{KKansatz}{
  ds_{str}^2 &= G_{\mu\nu} dx^\mu dx^\nu - e^{2\varphi_1} dt^2 +
   e^{2\varphi_2} dx_2^2 + e^{2\varphi_3} d\Omega_6^2  \cr
  B_2 &= b_2 + b_1 \wedge dx^2 + \tilde{b}_1 \wedge dt + 
   \tilde{b}_0 dt \wedge dx^2  \cr
   &H_3 = h_2 \wedge dx^2 + \tilde{h}_2 \wedge dt + 
     \tilde{h}_1 \wedge dt \wedge dx^2  \cr
  A_1 &= a_1 + a_0 dx^2 + \tilde{a}_0 dt  \cr
   &F_2 = f_2 + f_1 \wedge dx^2 + \tilde{f}_1 \wedge dt  \cr
  A_3 &= c_2 \wedge dx^2 + \tilde{c}_2 \wedge dt + 
   \tilde{c}_1 \wedge dt \wedge dx^2  \cr
   &F_4 = \tilde{f}_2 \wedge dt \wedge dx^2 \,,
 }
where $h_2 = db_1$, $\tilde{h}_2 = d\tilde{b}_1$, and so forth.  Forms such as $h_2$ and $b_1$ are constructed from the $dx^\alpha$, where $x^\alpha$ runs over $(x^1,r)$.

Plugging \KKansatz\ into the ten-dimensional action in \DzeroDtwo, one obtains a two-dimensional action that is the sum of two terms:
 \eqn{KKaction}{
  S &= \int d^2 x \, (L_{NS} + L_R)  \cr
  G^{-1/2} L_{NS} &= e^{-2\Phi} \Bigg( R + 
    4(\partial\Phi)^2 - 
   (\partial\varphi_1)^2 - (\partial\varphi_2)^2 -
   6 (\partial\varphi_3)^2 + 30 e^{-2\varphi_3}  \cr
    &\qquad\qquad{} - 
   {1 \over 2} e^{-2\varphi_2} h_2^2 + 
   {1 \over 2} e^{-2\varphi_1} \tilde{h}_2^2 + 
   {1 \over 2} e^{-2\varphi_1 - 2\varphi_2} \tilde{h}_1^2 \Bigg)  \cr
  G^{-1/2} L_R &= 
   -{1 \over 2} e^{\varphi_1+\varphi_2+6\varphi_3} f_2^2 
   -{1 \over 2} e^{\varphi_1-\varphi_2+6\varphi_3} f_1^2 +
   {1 \over 2} e^{-\varphi_1+\varphi_2+6\varphi_3} \tilde{f}_1^2 
    \cr&\qquad{} + 
   {1 \over 2} e^{-\varphi_1-\varphi_2+6\varphi_3}
    (\tilde{f}_2 - a_0 \tilde{h}_2 + \tilde{a}_0 h_2 + 
      a_1 \wedge \tilde{h}_1)^2 \,.
 }
Here $G = -\det G_{\alpha\beta}$ and $R$ refer to the two-dimensional string metric, and the two-dimensional dilaton is
 \eqn{twoDdilaton}{
  \Phi = 
   \phi - {1 \over 2} (\varphi_1 + \varphi_2 + 6 \varphi_3) \,.
 }
The Wess-Zumino term in the action in \DzeroDtwo\ vanishes for the ansatz \KKansatz.

The fields in \KKaction\ include four gauge fields: $b_1$, $\tilde{b}_1$, $a_1$, and $\tilde{c}_1$.  These fields are non-dynamical: they can be eliminated in favor of constants of the motion which are roughly their conjugate variables.  To illustrate this process of elimination, consider the following simple example:
 \eqn{SimpleGauge}{
  S = \int {\cal L} \,, \qquad
  {\cal L} = {1 \over 2} e^\varphi (f_2 + q_2) \wedge * (f_2 + q_2) + 
   q_0 f_2 \,,
 }
where $\varphi$, $q_2$, and $q_0$ depend arbitrarily on other fields.  The manipulations for eliminating $f_2$ are as follows:
 \eqn{EliminateGauge}{
  \delta {\cal L} &= e^\varphi d\delta a_1 \wedge * (f_2 + q_2) + 
   q_0 d\delta a_1 = \pi_a d\delta a_1  \cr
  \pi_a &\equiv e^\varphi *(f_2 + q_2) + q_0  \cr
  \hat{\cal L} &\equiv {\cal L} - \pi_a da_1 = 
   -{1 \over 2} e^{-\varphi} * (\pi_a - q_0)^2 + 
    (\pi_a - q_0) q_2 \,.
 }
The ``reduced lagrangian density'' $\hat{\cal L}$ is the analog of the Routhian in classical mechanics.  If one adds to the original action terms independent of $a_1$ and its derivatives, but depending on other fields which $q_0$ and $q_2$ may also depend on, then the equations of motion for these other fields may be obtained by varying the reduced lagrangian.

Starting with the lagrangian density for the form fields,
 \eqn{Lforms}{
  {\cal L}_f &= -{1 \over 2} 
    e^{-2\Phi-2\varphi_2} h_2 \wedge *h_2 + 
   {1 \over 2} e^{-2\Phi-2\varphi_1} \tilde{h}_2 \wedge * h_2 + 
   {1 \over 2} e^{-2\Phi-2\varphi_1-2\varphi_2} 
    \tilde{h}_1 \wedge *\tilde{h}_1  \cr&\qquad{} - 
   {1 \over 2} e^{\varphi_1+\varphi_2+6\varphi_3} f_2 \wedge *f_2 -
   {1 \over 2} e^{\varphi_1-\varphi_2+6\varphi_3} f_1 \wedge *f_1 +
   {1 \over 2} e^{-\varphi_1+\varphi_2+6\varphi_3} 
    \tilde{f}_1 \wedge * \tilde{f}_1
     \cr&\qquad{} +
   {1 \over 2} e^{-\varphi_1-\varphi_2+6\varphi_3} 
    (\tilde{f}_2 - a_0 \tilde{h}_2 + \tilde{a}_0 h_2 +
      a_1 \wedge \tilde{h}_1) \wedge
    * (\tilde{f}_2 - a_0 \tilde{h}_2 + \tilde{a}_0 h_2 +
      a_1 \wedge \tilde{h}_1) \,,
 }
One can eliminate first $\tilde{c}_1$, then $a_1$, then $b_1$ and $\tilde{b}_1$ in favor of the following constants:
 \eqn{piDefs}{
  \tilde\pi_c &= e^{-\varphi_1-\varphi_2+6\varphi_3} *(\tilde{f}_2 - 
    a_0 \tilde{h}_2 + \tilde{a}_0 h_2 + a_1 \wedge \tilde{h}_1)  \cr
  \pi_a &= -e^{\varphi_1+\varphi_2+6\varphi_3} *f_2 + 
    \tilde\pi_c \tilde{b}_0  \cr
  \pi_b &= -*e^{-2\Phi-2\varphi_2} h_2 + 
   \tilde\pi_c \tilde{a}_0  \cr
  \tilde\pi_b &= *e^{-2\Phi-2\varphi_1} \tilde{h}_2 -
   \tilde\pi_c a_0 \,.
 }
The reduced lagrangian density is
 \eqn{reducedL}{
  \hat{\cal L} &= 
   {1 \over 2} e^{2\Phi+2\varphi_2} 
    *(\pi_b - \tilde\pi_c \tilde{a}_0)^2
   -{1 \over 2} e^{2\Phi+2\varphi_1} 
    *(\tilde\pi_b + \tilde\pi_c a_0)^2
   +{1 \over 2} e^{-\varphi_1-\varphi_2-6\varphi_3} 
    *(\pi_a - \tilde\pi_c \tilde{b}_0)^2
   \cr&\qquad{} -{1 \over 2} e^{\varphi_1+\varphi_2-6\varphi_3}
     \tilde\pi_c^2
   +{1 \over 2} e^{-2\Phi-2\varphi_1-2\varphi_2} 
     \tilde{h}_1 \wedge * \tilde{h}_1
   -{1 \over 2} e^{\varphi_1-\varphi_2+6\varphi_3} f_1 \wedge *f_1
   \cr&\qquad{}
   +{1 \over 2} e^{-\varphi_1+\varphi_2+6\varphi_3} 
     \tilde{f}_1 \wedge *\tilde{f}_1 \,.
 }
The total reduced lagrangian (restoring the first line of $L_{NS}$ in \KKaction) is
 \eqn{RL}{
  G^{-1/2} L &= e^{-2\Phi} \Bigg( R + 4(\partial\Phi)^2 - 
   (\partial\varphi_1)^2 - (\partial\varphi_2)^2 -
   6 (\partial\varphi_3)^2 + 
   {1 \over 2} e^{-2\varphi_1 - 2\varphi_2} 
    (\partial\tilde{b}_0)^2 \Bigg)  
   \cr&\qquad{} 
   -{1 \over 2} e^{\varphi_1-\varphi_2+6\varphi_3} (\partial a_0)^2 
   +{1 \over 2} e^{-\varphi_1+\varphi_2+6\varphi_3}
     (\partial\tilde{a}_0)^2 
   \cr&\qquad{} + 30 e^{-2\Phi-2\varphi_3}
   +{1 \over 2} e^{2\Phi+2\varphi_2} 
     (\pi_b - \tilde\pi_c \tilde{a}_0)^2
   -{1 \over 2} e^{2\Phi+2\varphi_1} 
     (\tilde\pi_b + \tilde\pi_c a_0)^2
   \cr&\qquad{} 
   +{1 \over 2} e^{-\varphi_1-\varphi_2-6\varphi_3}
     (\pi_a + \tilde\pi_c \tilde{b}_0)^2
   -{1 \over 2} e^{\varphi_1+\varphi_2-6\varphi_3} \tilde\pi_c^2
    \,.
 }
Clearly, this lagrangian describes two-dimensional dilaton gravity coupled to the seven scalars $(\Phi,\varphi_1,$ $\varphi_2,\varphi_3,\tilde{b}_0,a_0,\tilde{a}_0)$, all of which participate in the potential energy function.

From \DzeroDtwo\ and \KKansatz\ one can easily read off the fields involved in the background solution:
 \eqn{background}{
  ds_{2,str}^2 &= {D \over \sqrt{H}} dx_1^2 + {\sqrt{H} \over h} dr^2
   \cr
  e^{2\varphi_1} &= {h \over \sqrt{H}} \qquad
   e^{2\varphi_2} = {D \over \sqrt{H}} \qquad
   e^{2\varphi_3} = \sqrt{H} r^2 \qquad
   e^{2\Phi} = \sqrt{D \over hH} {1 \over r^6}  \cr
  \tilde{a}_0 &= \coth\alpha \sin\theta
    \left( 1 - {1 \over H} \right) \qquad
  \tilde{c}_1 = \coth\alpha \sec\theta 
    \left( {D \over H} - 1 \right) dx   \cr
  b_1 &= \tan\theta \left( 1 - {D \over H} \right) dx^1 \,.
 }
The other fields in the action \KKaction\ vanish.  One easily obtains
 \eqn{GotPis}{\seqalign{\span\TL & \span\TR\qquad & \span\TL & \span\TR}{
  \tilde\pi_c &= -5r_0^5 \cosh\alpha \sinh\alpha \cos\theta &
  \pi_a &= 0  \cr
  \pi_b &= -5r_0^5 \sinh^2\alpha \cos\theta \sin\theta &
  \tilde\pi_b &= 0 \,.
 }}
It is straightforward though somewhat tedious to show that \background\ satisfies the equations of motion derived from \RL.

In obtaining the linearized perturbation equations, it is efficient to introduce a conformal factor: send $G_{\alpha\beta} \to e^{2\sigma} G_{\alpha\beta}$, so that the total reduced lagrangian becomes
 \eqn{RLagain}{
  G^{-1/2} L &= e^{-2\Phi} \Bigg( R - 
   4 \partial\sigma \partial\Phi 
   +4(\partial\Phi)^2 - 
   (\partial\varphi_1)^2 - (\partial\varphi_2)^2 -
   6 (\partial\varphi_3)^2 + 
   {1 \over 2} e^{-2\varphi_1 - 2\varphi_2} 
    (\partial\tilde{b}_0)^2 \Bigg)  
   \cr&\qquad{} 
   -{1 \over 2} e^{2\sigma+\varphi_1-\varphi_2+6\varphi_3} 
    (\partial a_0)^2 
   +{1 \over 2} e^{2\sigma-\varphi_1+\varphi_2+6\varphi_3}
     (\partial\tilde{a}_0)^2 
   \cr&\qquad{} + 30 e^{2\sigma-2\Phi-2\varphi_3}
   +{1 \over 2} e^{2\sigma+2\Phi+2\varphi_2} 
     (\pi_b - \tilde\pi_c \tilde{a}_0)^2
   -{1 \over 2} e^{2\sigma+2\Phi+2\varphi_1} 
     (\tilde\pi_b + \tilde\pi_c a_0)^2
   \cr&\qquad{} 
   +{1 \over 2} e^{2\sigma-\varphi_1-\varphi_2-6\varphi_3}
     (\pi_a + \tilde\pi_c \tilde{b}_0)^2
   -{1 \over 2} e^{2\sigma+\varphi_1+\varphi_2-6\varphi_3}
     \tilde\pi_c^2
    \,.
 }
Eight second order equations of motion are obtained by varying with respect to the eight scalars $\sigma$, $\Phi$, $\varphi_1$, $\varphi_2$, $\varphi_3$, $\tilde{b}_0$, $a_0$, and $\tilde{a}_0$.  It is not helpful to write out these equations in detail here.  One also has the Einstein equations, obtained from varying with respect to the metric.  If we compress \RLagain\ to the notation
 \eqn{RLoneMoreTime}{
  G^{-1/2} L &= e^{-2\Phi} (R + 4(\partial\Phi)^2) -
   {1 \over 2} {\cal G}_{ab} \partial\phi^a \partial\phi^b - 
    V(\phi) \,,
 }
where $\phi^a$ includes all eight scalars, then the Einstein equations read
 \eqn{EEqs}{
  e^{-2\Phi} &\left( R_{\mu\nu} - {1 \over 2} R G_{\mu\nu} + 
    2 G_{\mu\nu} (\partial\Phi)^2 + 2 \nabla_\mu \partial_\nu \Phi - 
    2 G_{\mu\nu} \nabla^2 \Phi \right)  \cr
   &= {1 \over 2} {\cal G}_{ab} 
     \partial_\mu\phi^a \partial_\nu\phi^b - 
    {1 \over 4} G_{\mu\nu} {\cal G}_{ab} 
     \partial\phi^a \partial\phi^b - 
    {1 \over 2} G_{\mu\nu} V(\phi) \,.
 }
The Einstein tensor $R_{\mu\nu} - {1 \over 2} R G_{\mu\nu}$ vanishes because we're working in two dimensions.  One combination of the Einstein equations (the trace) is equivalent to the equation of motion one gets by varying $\sigma$.  The other two independent combinations can be viewed as gauge conditions for the gauge choice of writing the perturbed metric as a conformal factor times the original one.

\subsection{Setting up a well-defined initial value problem}
\label{SETUP}

Because our interest is in a stationary mode with some wave-number $k$, let's restrict the ansatz by letting each of the eight scalars take the form
 \eqn{ScalarAnsatz}{
  \phi = \phi_0(r) + \delta\phi(r) \cos kx \,,
 }
where $\phi$ is one of the eight scalars and $\phi_0(r)$ is the background solution.  Terms proportional to $\sin kx$ turn out to be unnecessary: they decouple from the $\cos kx$ perturbations at first order.  Plugging \ScalarAnsatz\ into the equations of motion and constraints and expanding to linear order in the $\delta\phi(r)$ leads to linear ordinary differential equations.  These linearized equations comprise eight order equations of motion (obtained by varying the two-dimensional lagrangian \RLagain) and two constraints (obtained from the Einstein equation) in the eight variables $\sigma$, $\Phi$, $\varphi_1$, $\varphi_2$, $\varphi_3$, $\tilde{b}_0$, $\tilde{a}_0$, and $a_0$.  The form of these ten equations is still remarkably complicated and will not be reproduced here.

The ten equations described incorporate some redundancy: six of the equations of motion plus the two constraints imply the remaining two equations of motion.  One can therefore drop the equations of motion for $\delta\sigma$ and $\delta\Phi$.  It is found by direct computation that the equations of motion for $\delta a_0$ and $\delta\tilde{b}_0$ decouple from the others, and that these two fields may be set uniformly to zero.  What remains is six coupled equations in the six variables 
 \eqn{FinalVariables}{
  q_i = (\delta\sigma,\delta\Phi,\delta\varphi_1,\delta\varphi_2,
   \delta\varphi_3,\delta a_0) \,.
 }
They comprise the equations of motion for $\delta\varphi_1$, $\delta\varphi_2$, $\delta\varphi_3$, and $\delta a_0$, plus the two constraints from the Einstein equations.  These linear equations have terms multiplying $q_i(r)$, $q_i'(r)$, and $q_i''(r)$, except that $\delta\sigma''(r)$ and $\delta\Phi''(r)$ do not appear.  Let us call these six differential equations the perturbation equations.

Boundary conditions at infinity are simple: one should require that the $q_i$ are normalizable.  For $k \neq 0$, the behavior for large $r$ is $q_i \sim e^{\gamma_i k r}$ for some constants $\gamma_i$.  A normalizable solution is one in which all the $\gamma_i$ are negative.

Boundary conditions at the horizon are also conceptually simple: one should require that the perturbed metric and matter fields are smooth at the horizon in ten dimensions.  This requirement becomes rather obscure in our two-dimensional language.  So I implement boundary conditions in a different way, which is somewhat complicated to state, but which I expect is equivalent.  Start with the eight non-trivial equations obtained by setting $\delta a_0 = \delta\tilde{b}=0$ in the original ten equations described below \ScalarAnsatz.  Then set $r=r_0$.  It happens that all terms involving $q_i''(r_0)$ drop out: the second derivatives are multiplied by functions which vanish at $r=r_0$.  So the result is eight linear algebraic relations on the twelve quantities $(q_i(r_0),q_i'(r_0))$.  Boundary conditions at $r=r_0$ on differential equations which do not involve $\delta\sigma''(r)$ and $\delta\Phi''(r)$ should not themselves depend on $\delta\sigma'(r_0)$ or $\delta\Phi'(r_0)$.  Imposing this limitation, one goes from eight relations to five, now involving the ten quantities
 \eqn{FinalCauchyVariables}{
 \delta\sigma,\delta\Phi,
   \delta\varphi_1,\delta\varphi_2,
   \delta\varphi_3,\delta a_0,
 \delta\varphi_1',\delta\varphi_2',
   \delta\varphi_3',\delta a_0' \,.
 }
Of these conditions, the simplest are
 \eqn{TwoBCS}{
  \delta\tilde{a}_0 = 0 \,,\qquad \delta\varphi_1 = \delta\sigma \,.
 }
The first of these states that the D0-brane voltage must remain constant across the horizon after the perturbation.  The second states that temperature must remain constant.

Clearly, for any point $r>r_0$, Cauchy data for the perturbation equations consists of a value for the ten quantities in \FinalCauchyVariables.  But because of the reduction of order at $r=r_0$, one cannot simply specify Cauchy data right at the horizon!  Instead, one takes the five relations on the quantities \FinalCauchyVariables, obtained by setting $r=r_0$, and imposes them at $r=r_0+\epsilon$ for some small $\epsilon$: these conditions are the stand-ins for the true horizon boundary conditions, and I believe they become equivalent in the limit $\epsilon \to 0$.

To specify Cauchy data at $r=r_0+\epsilon$, one must set values for five of the quantities \FinalCauchyVariables\ in such a way that the other five can be determined from the boundary conditions described in the previous paragraph.  One appropriate choice of these five quantities is $\delta\sigma, \delta\Phi, \delta\varphi_1', \delta\varphi_2, \delta\varphi_3$.  Of course, one must also specify $r_0$, $\epsilon$, and $k$.  As usual with classical equations, there is a single scaling symmetry corresponding to rescaling the Planck length.  This allows one to set $r_0=1$.  Because the perturbation equations are linear, there is another scaling symmetry: $q_i(r) \to \lambda q_i(r)$ for arbitrary $\lambda$.  Let us use this symmetry to set $\delta\sigma(r_0+\epsilon)=1$.  Let us also regard $\epsilon$ as a fixed quantity.  Then the quantities to be varied are
 \eqn{FinalInputs}{
  k,\delta\Phi,\delta\varphi',\delta\varphi_2,\delta\varphi_3 \,,
 }
where the functions are all evaluated at $r=r_0+\epsilon$.

The strategy to find a normalizable perturbation is to vary the five quantities \FinalInputs, integrate the perturbation equations, and check normalizability at some large $r$.  The equations at large $r$ have five normalizable and five non-normalizable solutions, so with the five variables \FinalInputs\ in hand one has just enough freedom to get rid of the non-normalizable solutions.
  
The computations summarized in this section are entirely analytical, and they are in large part available in the form of a Mathematica notebook \cite{notebook}.

\subsection{Results of numerics}
\label{RESULTS}

To recap the previous section: having set $r_0=1$ and fixed a small $\epsilon$, one wants to vary the five quantities \FinalInputs\ so as to get a solution to the perturbation equations which is normalizable for large $r$.  In practice, once Cauchy data is specified, I integrated the perturbation equations out to a certain radius $r_f$, and then the test of normalizability was to evaluate the sum of the $L^2$ norms squared of the six functions $q_i(r)$ over the rightmost fifth of the integration region of the perturbation equations, minus the rightmost hundredth---that is, roughly for $r \in (0.8 r_f, 0.99 r_f)$ when $r_f \gg 1$.  Let us call this region $I$.  Then the quantity to be minimized is
 \eqn{TailNorm}{
  T = \int_I dr \sum_i q_i(r)^2 \,.
 } 
Such an integrated test avoids an unpleasant possibility that arises if one simply makes $q_i(r_f)=0$ the requirement at large $r$: zeroes of solutions to the perturbation equations can arise by making several growing exponentials cancel.  Also, it's awkward to implement a test condition of the form $q_i(r_f)=0$ because there are six functions and only five variables \FinalInputs\ to adjust.

One is left with the problem of minimizing the everywhere-positive function $T$, which must be numerically computed for any choice of the five variables \FinalInputs.  If the CSC is correct, the absolute minimum of $T$ is a lot smaller when $(\alpha,\theta)$ are in the thermodynamically unstable region.  But it's hard to find the absolute minimum of a function of five real variables which is computationally expensive to evaluate.  I employed the Mathematica function {\tt NMinimize}, which combines several standard approaches to minimization.  It samples on the order of $10^4$ points, which is actually rather few when one considers that $6^5 = 7776$.  I used $\epsilon = 1/20$ and $r_f = 5$ and considered $16$ different points in the $(\alpha,\theta)$ plane, shown in figure~\ref{figA}.

In performing the numerical integrations, it was noted that when $k=0$, there are static perturbations of the black hole which are regular at infinity but not normalizable: the $q_i(r)$ approach constant values.  These perturbations are nearby supergravity solutions with slightly different charges or mass.  This creates a difficulty in determining the correct normalizable solutions: the function $T$ described in \TailNorm\ has a relatively broad, shallow minimum around $k=0$, and numerical minimization of $T$ is usually drawn to it even when the absolute minimum is at $k \neq 0$.  One solution was to tell {\tt NMinimize} never to try a value of $k$ less than some arbitrary lower limit (for example, $0.3$).  Another was to minimize not $T$ itself, but rather $\tilde{T} \equiv e^{-1.4 k r_f} T$.  The choice of prefactor was arranged specially to eliminate the broad shallow minimum without causing $\tilde{T}$ to vanish exponentially at large $k$: indeed, $\tilde{T}$ still has weak exponential growth at large $k$.  This second method proved quite effective at finding the normalizable solution when the CSC says there should be one.  However, for reasons I do not understand, it seemed less good than unrestricted minimization of $T$ itself at finding the least divergent solution when there is no normalizable mode.  Perhaps we should not be too surprised: if numerics are optimized to do one thing well, they may do another thing less well.

We can also understand at this point why it is difficult (though not impossible) to test the hypothesis that the wavelength of the stationary perturbation diverges as one passes to the NCFT limit.  Diverging wavelength means small $k$, and for small $k$ it becomes harder to distinguish between normalizable solutions and solutions which are regular at infinity but not normalizable.

\begin{figure}
  \centerline{\includegraphics[width=3in]{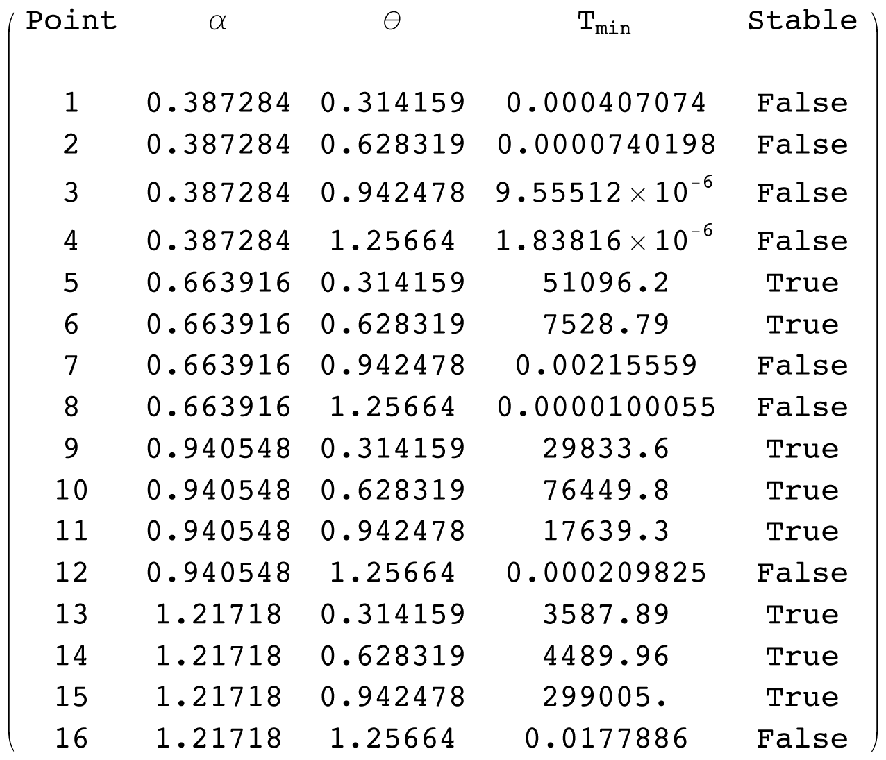}\hfil
   \lower7pt\hbox{\includegraphics[width=2.7in]{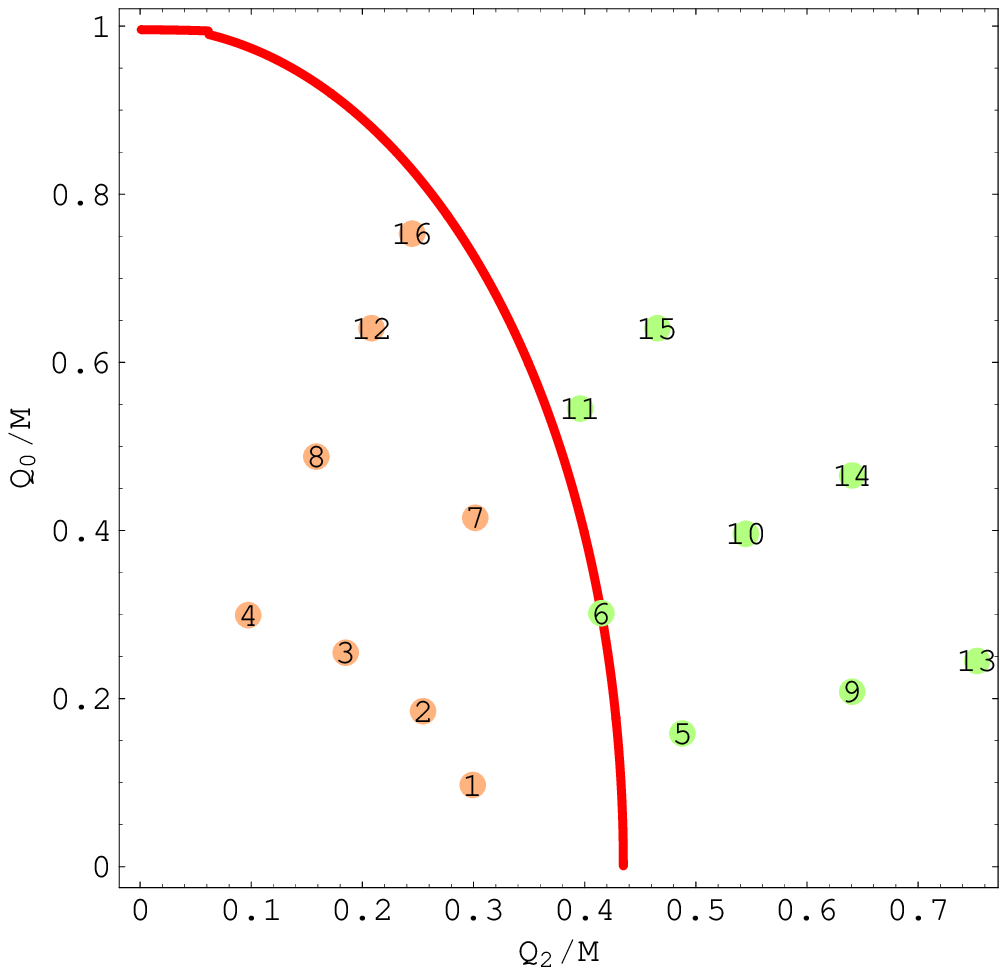}}}
  \caption{Left: the minimum of $\tilde{T}$ for various values of $\alpha$ and $\theta$.  In the final column, ``False'' means that a normalizable solution should exist according to the CSC, and ``True'' means that it shouldn't.  Right: the position of these $16$ points in the $Q_2/M$, $Q_0/M$ plane.  A red dot indicates that a normalizable mode was found, and a green dot indicates that it wasn't.}\label{figB}
 \end{figure}

\begin{figure}
  \centerline{\includegraphics[width=3.5in]{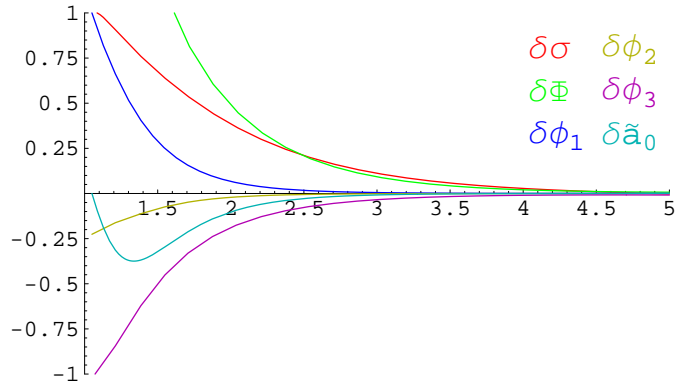}}
 \caption{The normalizable mode for the point labeled 12 in figure~\ref{figB}.}\label{figC}
 \end{figure}

An overview of the results of numerics is presented in figure~\ref{figB}.
Figure~\ref{figC} shows a plot of the six scalars involved in the perturbation for a particular choice of $\alpha$ and $\theta$ where an instability exists.
The numerical results described represent on the order of 100 hours of CPU time on a $2.4\,{\rm GHz}$ Xeon processor (that is, a fast PC by 2004 standards).

\section{Conclusions}
\label{CONCLUDE}

The D2-D0 bound state obviously must exhibit a Gregory-Laflamme instability sufficiently far from extremality.  The correlated stability conjecture predicts a particular threshhold at which the GL instability sets in: it is where local thermodynamic stability, defined in terms of the Hessian matrix of susceptibilities, is lost.  This threshhold is described by a curve in the plane of $Q_2/M$ and $Q_0/M$, shown in figure~\ref{figA}.  $Q_2/M$ is the mass fraction of the D2-branes; $Q_0/M$ is the mass fraction of the D0-branes; and the remaining mass fraction is from the non-extremality.

The stability threshold predicted by the CSC passes from a finitely non-extremal D2-brane with no D0-branes to a near-extremal bound state whose mass comes mostly from D0-branes.  The dynamics of this near-extremal bound state is supposed to be non-commutative field theory at finite temperature, but the Gregory-Laflamme instability occurs for a temperature that is low compared to the intrinsic scale $1/\sqrt{\vartheta}$ of the NCFT.  I conjecture that the wavelength of this instability becomes very long in the NCFT limit, avoiding a contradiction with decoupling arguments.  But it is worth noting that for a specified D2-D0 configuration that is close to the NCFT limit, the supergravity approximation (according to the CSC) predicts the existence of an instability at very low temperatures that makes the non-commutativity parameter depend on space (and probably time).  This instability is difficult to see in a weakly coupled description.

To verify the predictions of the CSC, we have looked numerically for a normalizable, stationary perturbation of the D2-D0 bound state for various values of $Q_2/M$ and $Q_0/M$.  It helped first to reduce the problem to two dimensions parametrized by $r$ and $x^1$ (the direction in which the inhomogeneity develops).  Two points in the numerical analysis deserve some scrutiny:
 \begin{enumerate}
  \item Boundary conditions at the horizon were implemented in a slightly ad hoc fashion, motivated by the realization that the horizon is a regular singular point of the equations of motion when expressed in terms of a standard Schwarzschild-like radial variable.  But the simplest of these boundary conditions are physically correct, and as a whole they clearly pick out finite perturbations at the horizon; so I have little doubt that the boundary conditions are indeed correct.
  \item The numerics were ``trained'' to avoid a broad shallow local minimum for small wave-number so as to more efficiently find the global minimum (when an instability exists) at finite wave-number.  It is a bit disappointing that optimizing the numerical routines in this way made them worse at finding the least divergent perturbation in the regime where the CSC predicts no instability.  However, it would be all but inconceivable to find that an instability exists when the CSC says it shouldn't: such a situation would amount to a violation of the Second Law of Thermodynamics as formulated in classical supergravity.  So this optimization issue doesn't loom very large in the final analysis.
 \end{enumerate}
The bottom line from the numerical study is that the predictions of the CSC are spectacularly confirmed, with a clean separation of stable and unstable points and scalar profiles which exhibit the expected exponential tails as well as some interesting structure near the horizon (figure~\ref{figC}).

A variety of other brane bound states can be studied by similar methods.  It is fairly straightforward to make predictions of the stability boundary using the CSC, and a typical behavior is for the stability boundary to pass into the near-extremal regime in a limit where most of the mass comes from branes that are delocalized in some spatial dimension.  Only, for D4-D0, where the binding energy is zero, one naturally expects a GL instability for all values of D0-brane charge and non-extremality.  One may also consider adding angular momentum: it is an additional thermodynamic quantity which can be locally redistributed, so it can participate in GL instabilities.\footnote{Thermodynamic instabilities of D-branes with angular momentum were first studied in \cite{gSpin}, and the stability boundary for general spinning D3-branes was found in \cite{cgTwo}.}  Studying the linearized equations of motions around brane bound states is in general somewhat complicated, but we are fairly confident that the CSC would be confirmed by such analysis in all the cases we have described here.\footnote{A possible counter-example to the CSC was suggested in \cite{gyrating} based on solutions where angular momentum can be carried by a string in more than one way---either as a property of the horizon or through mechanical gyration.  Because the arguments in \cite{gyrating} are almost wholly thermodynamic, I would suggest that if dynamical calculations indeed show an unstable mode in linear perturbation theory, the separation of angular momentum into two distinguishable components may be precisely the refinement of thermodynamic arguments necessary to make the CSC work in this case.}  I hope to report on these issues in the future; work on some of them is already under way \cite{gFriessUnfinished}.

\section*{Acknowledgements}

I thank G.~Horowitz and D.~Marolf for useful discussions.  This work was supported in part by the Department of Energy under Grant No.\ DE-FG02-91ER40671, and by the Sloan Foundation.

\bibliographystyle{ssg}
\bibliography{smear}

\end{document}